\documentstyle[draft,psfig]{mn}

\begin{document}

\title[CMB anisotropy: Non-Gaussianity]{CMB anisotropy: 
deviations from Gaussianity due
to non-linear gravity}
\author[Aliaga et al.]{A.M. Aliaga$^{1}$, V. Quilis$^{2}$, 
J.V. Arnau$^{3}$
\& D. S\'aez$^{1,4}$\\ $^{1}$ Departamento de Astronom\'{\i}a y 
Astrof\'{\i}sica, Universidad de Valencia.
46100 Burjassot, Valencia, Spain\\
$^{2}$Department of Physics, Durham University, South Road,
Durham, DH1 3LE, UK\\
$^{3}$Departamento de Matem\'atica Aplicada,
Universidad de Valencia.
46100 Burjassot, Valencia, Spain\\
$^{4}$email:diego.saez@uv.es\\ } \maketitle

\begin{abstract}
Non-linear evolution of cosmological 
energy density fluctuations triggers 
deviations from Gaussianity in
the temperature
distribution of the cosmic microwave background.
A method to estimate these deviations is proposed.
N-body simulations -- in a $\Lambda$CDM cosmology -- 
are used to 
simulate the strongly non-linear evolution of 
cosmological structures. It is proved that these simulations can
be combined with the potential approximation to calculate 
the statistical moments
of the CMB anisotropies produced by non-linear gravity.
Some of these moments are computed and the resulting values 
are different from 
those corresponding to Gaussianity. 
\end{abstract}


\section{INTRODUCTION}

Many experiments have been designed to analyze the Cosmic
Microwave Background (CMB) anisotropy. The signal to be 
detected is the superposition of 
various components. In the inflationary models considered 
here, the components generated during linear
evolution (Sachs-Wolfe, Doppler, and so on) are dominant and Gaussian;
nevertheless, there are subdominant non Gaussian components as 
those due to
non-linear gravity (Rees-Sciama), lens distortions, and contaminant
foregrounds. The 
characterization of all these subdominant components, including
deviations with respect to Gaussianity, is necessary for
data analysis in experiments as MAP and PLANCK.
The Rees-Sciama effect was studied,
using N-body simulations, by 
Seljak (1996) and
Tuluie, Laguna \& Anninos (1996); nevertheless,  
deviations
with respect to Gaussianity were not considered at all
by these authors.
The third order moment
of the Rees-Sciama effect was estimated,
using second order 
perturbation theory
(not a fully 
non-linear method describing strongly 
non-linear evolution),
by 
Mollerach et al. (1995) and Munshi, Souradeep \& Starobinski (1995).
Here,
we are mainly concerned with the deviations with respect to 
Gaussianity produced by 
fully non-linear gravity (Rees-Sciama effect). 
A new general numerical method specially
designed to estimate these deviations 
is proposed. It is based on N-body simulations.
Using this method, some statistical moments
are estimated for the first time.

Quantities 
$\Omega$ and $\Omega_m$ are defined as follows:
$\Omega = \Omega_{b} + \Omega_{d} + \Omega_{\Lambda}$ and
$\Omega_{m} = \Omega_{b} + \Omega_{d}$, where
$\Omega_{b}$, $\Omega_{d}$ and  $\Omega_{\Lambda}$ stand for the
density parameters corresponding to baryons,  dark matter, and  
vacuum, respectively.
Quantity  h is the reduced Hubble constant. 
We work in the framework of 
a standard inflationary (flat) model with cold dark matter,
having $h=0.65$,
$\Omega_{b}=0.05$, $\Omega_{d}=0.25$, and $\Omega_{\Lambda}=0.7$.
The power spectrum has been normalized with $\sigma_8=0.93$ according 
with Eke et al.(1996), who estimated $\sigma_8$ to get
cluster abundances compatible with observations.
In such a model, scalar
energy density fluctuations are initially Gaussian, and
they remain 
Gaussian during linear evolution; nevertheless, 
deviations with respect to Gaussianity arise 
in the non-linear evolution (where Fourier modes mix).

In this paper, we are concerned with the CMB temperature 
distribution.
This field  
can be developed in spherical harmonics
($\Delta T/T = \sum_{\ell=0}^{\infty}
\sum_{m = -\ell}^{+\ell} a_{\ell m} Y_{\ell m}$). 
Taking into account the cosmological principle,
the CMB temperature is assumed to be 
a realization 
of a homogeneous and isotropic statistical field and,
as a consequence of this assumption, the $m$ dependence of
the bispectrum has  --in terms of the Wigner-3j symbols--
the well known form : 
\begin{equation} \label{eq:ecu208}
\langle a_{\ell_{1} m_{1}} a_{\ell_{2} m_{2}} a_{\ell_{3} m_{3}}
\rangle = B_{\ell_{1} \ell_{2} \ell_{3}} \left(
\begin{array}{ccc}
\ell_{1} & \ell_{2} & \ell_{3} \\
m_{1} &  m_{2} & m_{3} 
\end{array}
\right) \ .
\label{bi}
\end{equation}
On account of this formula, we can 
take $m_{1} = m_{2} = m_{3} = 0$ to compute 
$B_{\ell_{1} \ell_{2} \ell_{3}}$ and, then, Eq. (\ref{bi})
allows us to calculate
$\langle a_{\ell_{1} m_{1}} a_{\ell_{2} m_{2}} a_{\ell_{3} m_{3}}
\rangle$ for other $m$ values. The $m$ dependence 
of the trispectrum is given explicitly by Hu (2001).

If non-linear evolution is not considered at all, namely,
under the assumption that the CMB anisotropy is only 
produced by linear inflationary
fluctuations (which are Gaussian as a result of the 
generation mechanism),               
the $a_{\ell m}$ coefficients appear to be statistically 
independent Gaussian quantities with zero means and 
variances $\langle \mid a_{\ell m} \mid^{2} \rangle
= C_{\ell}$.
In this situation (see Grishchuk \& Martin 1997),
all the odd moments (mean, bispectrum, and so on) vanish,
the spectrum is given by
$\langle  a_{\ell_{1} m_{1}}  a_{\ell_{2} m_{2}} \rangle = C_{\ell_{1}}
\delta_{\ell_{1} \ell_{2}} \delta_{m_{1},-m_{2}}$, and the 
higher order even moments can be written as functions of the
$C_{\ell}$ quantities , in
particular, the relation $\langle \mid a_{\ell m} \mid^{4} \rangle
= 3 C_{\ell}^{2}$ is satisfied.  
In order to test that non-Gaussianity develops during
non-linear evolution,
deviations with respect to this relation 
and also with respect to a vanishing
bispectrum are investigated.

Along this paper, 
units are chosen in such a way that the speed of light is $c=1$ and,
whatever quantity "$A$" may be, $A_{_{L}}$ and $A_{0}$ stand for
the $A$ values on 
the last scattering surface and at present time,
respectively.

\section{ANGULAR SCALES, BOXES, AND RESOLUTION}

We focus our attention on the anisotropy generated while the
CMB photons move through the gravitational field 
produced by non-linear cosmological structures.
It is worthwhile to distinguish two types of 
these structures: (1) mildly non-linear structures
having  spatial scales 
larger than those of the clusters (superclusters, voids,
Great Attractor-like structures), and (2)
strongly non-linear structures  with spatial scales either
smaller or equal to those of the clusters. 
In case (1), computations can                 
be performed by using second order 
perturbation theory as in 
Mollerach et al. 1995 and Munshi, Souradeep \& Starobinski 1995
or, perhaps,
the Zel'dovich approach and its generalizations;
nevertheless, 
case (2) requires numerical computations based on
N-body schemes. We focus our attention on 
strongly non-linear structures as clusters,  which could produce
significant deviations from gaussianity, in particular, 
during the process of violent relaxation. 
These structures started to be non-linear (to undergone 
violent relaxation)
at low redshifts of about $z \leq 30$  ($z \leq 3$).

In our flat $\Lambda$CDM model, the potential 
approximation 
(Mart\'{\i}nez-Gonz\'alez, Sanz \& Silk 1990, 1994; 
Sanz et al. 1996), can be used to get the following basic equation: 
\begin{equation}
\frac {\Delta T} {T} = 
- \frac {5}{3} \phi_{_{L}} - \vec {n} \cdot \vec {v}_{_{L}} - 
2 \int_{t_{_{L}}}^{t_{o}} \vec{\nabla} \phi \cdot d \vec{x}
\ , 
\label{efun}
\end{equation}  
and
\begin{equation}
\Delta \phi = 4 \pi G \delta a^{2} \rho_{_{B}}  \ , 
\label{lapla}
\end{equation}
where $\frac {\Delta T} {T}$
is the relative temperature variation 
--with respect to the background temperature-- along the  
direction $\vec {n}$, and 
the integral is to be computed, from emission at the last
scattering surface ($L$) to
observation ($O$), along the background null geodesics.
Symbols $x^{i}$ ($\vec {x}$),
$\phi$, $\vec {v}$, $\rho_{_{B}}$, $\delta$, $a$, $t$, $G$,
stand for the comoving coordinates, the peculiar gravitational potential,  
the peculiar velocity,
the background mass density, the density contrast, the scale factor,
the cosmological time, and the gravitational constant, respectively.
Since we are interested in a gravitational effect,
the subdominant baryonic component 
can be neglected and, consequently, the density contrast -- appearing in 
Eq. (\ref{lapla})--
can be evolved using N-body simulations. Our simulations start
at $z=50$ to be sure that all the period of non-linear evolution 
is taken into account. 
The Rees-Sciama effect produced by non-linear structures 
located at redshift $z \leq 50$ is given by the third term
of the r.h.s. of Eq. (\ref{efun}), which can be approximated  
as follows:
$\left[ \frac {\Delta T} {T} \right]_{_{I}} \simeq 
- 2 \int_{t_{50}}^{t_{o}} \vec{\nabla} \phi \cdot d \vec{x}
\simeq 2 \int_{t_{50}}^{t_{o}} \frac {\partial \phi (\vec{x},t)}
{\partial t} dt$,
where $t_{50}$ stands for the cosmological time at redshift $z=50$.
Since the anisotropy under consideration can be
calculated as an integral of $\partial \phi (\vec{x},t) /
\partial t$, the angular scales of this 
anisotropy are not those subtended by the density contrast 
itself, but those corresponding to                                 
the potential $\phi$ (see Fullana \& S\'aez 2000).

The remainder of this section is a qualitative discussion 
about angular scales and resolution. No accurate computations
are necessary and spherically symmetric structures can be considered.
According to the cosmological principle, 
any spherical cosmological structure is compensated 
at a certain distance or compensation
radius $r = \zeta$. The peculiar mass (excess or defect)
inside a sphere of radius $r$,  
hereafter $M(r)$, vanishes at $r \geq \zeta$ and,
taking into account  the 
relation 
\begin{equation} 
\frac {\partial \phi} {\partial r} \propto \frac {M(r)} {r^{2}} \ ,
\label{compen}
\end{equation} 
which follows from Eq. (\ref{lapla}) in the spherically symmetric case, 
the potential $\phi $ does not depend on $r$
for $r \geq \zeta$, furthermore, this constant potential
must vanishes in order to get a good asymptotic 
behaviour. This means that a given compensated structure
only produces a significant gravitational potential
in the region $0 < r < \zeta$; namely, inside a sphere
with diameter $2 \zeta$.
In the linear regime, the peculiar velocity produced by a
structure is $\vec {v} \propto \vec {\nabla} \phi$. Although 
this relation does not hold in the central non-linear region
of a cluster, it roughly holds in the outer regions, where
compensation takes place and the density contrast is low.
Hence, the peculiar velocity vanishes 
for $r \geq \zeta$, at the same place where the gravitational
potential vanishes by compensation.

If the clustering of clusters is neglected
-- a good approach in our qualitative analysis-- and
the present mean separation between neighbouring 
clusters is $\xi h^{-1}$ Mpc, then,
the mean distance, at redshift $z$, is $\xi /(1+z)h$ Mpc 
and, consequently, given two similar neighbouring clusters,
each of them should become roughly
compensated at the centre of the segment joining the 
cluster centres (where peculiar velocity vanishes); 
therefore, the radius of the region where the 
potential of a given cluster --at redshift $z$-- is not negligible  
appears to be $\zeta = \xi /2(1+z)h$.
Let us now consider Virgo cluster, which is a standard cluster.
It is well known that the
peculiar velocity produced by the Virgo cluster 
on the Local Group ranges 
between $200$ and $300$ Km/s; hence, taking into account 
that the distance from the Local   
Group  to the cluster centre is $\sim 20$ Mpc
($\sim 13 h^{-1}$ Mpc), we can conclude that 
the Virgo cluster has not been fully compensated at 
a distance of $\sim 13 h^{-1}$ Mpc from its centre.
This strongly suggests that, effective compensation should take place 
between  $\zeta = 10 h^{-1}$ Mpc ($\xi=20$)
$\zeta = 20 h^{-1}$ Mpc ($\xi=40$); this second value of $\zeta$
seems a bit large, although
an admissible value in the case of 
Abell clusters, for which,
compensation should be produced at distances
greater than those corresponding to Virgo-like structures.

In a flat universe with cosmological constant,
the angle subtended by the distance $\xi /(1+z)h$ Mpc
placed at redshift $z$ is
\begin{equation} 
\Delta \theta = \frac {\xi} {3000f(z)}
\end{equation} 
where
\begin{equation} 
f(z)=\int_{0}^{z} \frac {d \eta} {[\Omega_{m0} (1+ \eta )^{3} +
\Omega_{\Lambda}]^{1/2}} \ ;
\end{equation}
using these formulae, we have calculated the $\ell$ value
corresponding to the angular scale $\Delta \theta$
($\ell= \pi / \Delta \theta $) as a
function of $z$. The results are displayed in 
Fig. 1 for
$\xi = 20$ (solid line), $\xi = 30$ (dotted line),
$\xi = 40$ (dashed line), and $\xi = 50$ (dashed-dotted line). 
In this Figure 
we see that, from $z \sim 0.5$ to $z \sim 5$, where clusters 
undergo violent relaxation and the non-linear gravitational anisotropy 
should be mainly produced, the $\ell $ values range 
in the interval (100,800). For $z<0.5$, clusters 
have relaxed and, for $z > 5$, they are at the first stages
of their non-linear evolution. 

\begin{figure}
\begin{centering}
\psfig{file=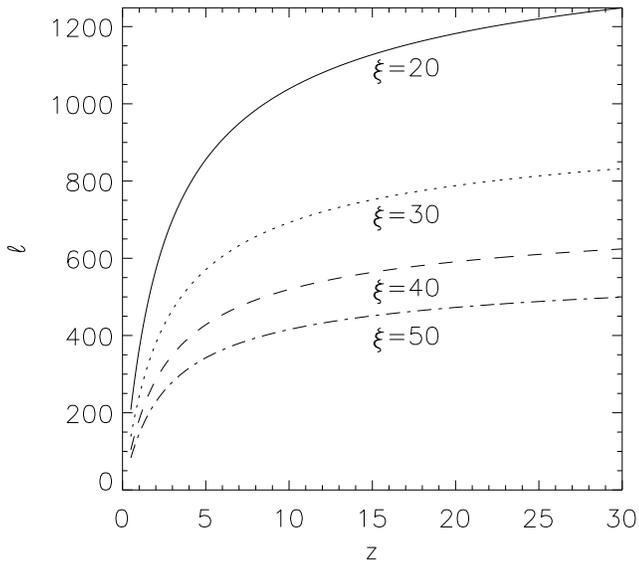,width=8.4cm}
\end{centering}
\caption{$\ell$ scale of the 
non-linear gravitational anisotropy
produced by non-linear structures as a function of 
their redshifts $z$. Four
values of the present scale $\xi $ --defined in the text-- 
are considered.}
\label{fig-lx-error}
\end{figure}

For the sake of briefness and clearness,
only the scale $\ell = 400$ (with $m=0$) has been considered in
our calculations. Let us now justify this choice
taking into account Fig. 1.
The method proposed in this paper 
might be sensitive to both the degree of
non-linearity and the speed of the $\phi$ evolution.
Strong non-linearities and large 
values of $\partial \phi / \partial t$ (violent 
relaxation) could lead to numerical problems. Non-linearity
requires good spatial resolution in the N-body simulation, 
and fast 
$\phi$ evolution requires appropriate time steps
in both the N-body simulation, and the numerical integrations 
along the line of sight (see below).
Fig. 1 shows that non-linear clusters 
of different sizes located between 
redshifts 0.5 and 5 (some of them undergoing violent relaxation) 
are contributing to $\ell = 400$, whereas only small clusters
located at $z \sim 5$ (well before violent relaxation)
are contributing to $\ell=800$; therefore, $\ell = 400$
seems to be a good choice in order to 
test our method taking into account 
both non-linear
evolution and violent relaxation.
Moreover, in the model under consideration,
the first acoustic peak is located at $\ell=200$ and,
around $\ell=400$, there is a minimum of the linear dominant 
Gaussian effects.

Cluster evolution is simulated in boxes having --at present time--
$128$ Mpc per edge with a resolution of
$\sim 1$ Mpc. 
This choice is a balance between the need of having a
reasonable number of clusters inside the box (order ten)
and the need of following the non-linear evolution 
of the gravitational potential.
The minimum scale we resolve, $\sim 1$ Mpc, 
is typically of the order of the cluster 
virial radius and, consequently, non-linear enough. 
The fine structure within
the core cluster is not resolved, but 
it does not seem to be necessary because the effect under 
consideration is produced by the gravitational potential,
which smoothes all the substructure on the scales
we are interested in. The effects produced
by substructures ($\xi << 20$ and strong clustering) 
should appear on angular scales 
corresponding to $\ell >> 400$, for which more resolution 
would be necessary.

N-body simulations are performed by using a standard
particle-mesh code (Hockney \& Eastwood 1988), which 
was used, tested, and described in more detail by Quilis,
Iba\~{n}ez \& S\'aez (1998).

We are interested in non-Gaussianity and, consequently,
our goal is not the computation of the angular power 
spectrum ($C_{\ell}$ coefficients) corresponding to 
the third term of Eq. (\ref{efun}). This spectrum was given 
by  Seljak (1996) for the Rees-Sciama effect, and by
Fullana \& S\'aez (2000) for the Integrated Sachs-Wolfe
effect. 
We are looking for the bispectrum,
trispectrum and so on. This fact 
facilitates our implementation of 
N-body simulations. In 
the case $\ell = 400$, according
to Fig. 1, cluster-like structures ($\xi < 40$) 
are contributing to
the Rees-Sciama effect inside the redshift interval (0.5,5),
whereas structures as superclusters, the Great Attractor,
and so on ($\xi > 50$), are 
contributing for $z > 10$, just when they were almost linear
objects; hence, 
these structures can only produce a very small almost Gaussian Integrated 
effect for
$\ell = 400$ (see Fullana \& S\'aez 2000), and no 
important deviation from Gaussianity are expected; hence, 
even if we partially include
superclusters and larger structures (small box size), we can get 
very accurate estimations of the deviations 
with respect to Gaussianity (associated to strong non-linearity).

Since we are mainly interested in the evolution of
the gravitational potential of
clusters and substructures and, moreover,
the chosen box is much larger than
the size of the region where the potential
of standard clusters is significant,
the Fourier transform can be performed in 
our box (as it is done in next Section); namely,  
no important artifacts are expected
as a result of the periodicity of an universe filled
by boxes identical to the chosen one.

\section{DEPARTURES FROM GAUSSIANITY}

The basic equations 
necessary for the numerical computation of
the bispectrum and other statistical moments are first derived
using the potential approximation and, then, 
a numerical method for the implementation of these equations --based on
N-body simulations-- is described and applied.
 
At present time, the observer is assumed to be 
located at a certain point with comoving 
spatial coordinates $x^{i}_{_{P}}$. 
The equations of the null geodesic passing by
point $x^{i}_{_{P}}$ are: 
\begin{equation} 
x^{i}=x^{i}_{_{P}} + \lambda (a) n^{i} \ ,
\label{ng}
\end{equation}
where the function $\lambda (a)$ can be 
written as follows:
\begin{equation} 
\lambda (a)= \kappa (a) H_{0}^{-1} \ ,               
\end{equation}     
with
\begin{equation} 
\kappa(a) = \int_{a}^{1} \frac {db} {(\Omega_{m0}b+
\Omega_{\Lambda} b^{4})^{1/2}} \ .
\end{equation}
Since we are interested in the deviations from Gaussianity 
produced by non-linear gravity, the
third term of the r.h.s. of Eq. (\ref{efun}) 
is calculated. Using Eq. (\ref{ng}), this term can be
written as follows:
\begin{equation} 
\left[ \frac {\Delta T} {T} \right]_{_{I}} (\vec{x}_{_{P}},\vec{n}) = 
\Delta (\vec{x}_{_{P}},\vec{n}) =
-2 \int_{t_{50}}^{t_{o}} (\vec{\nabla} \phi \cdot \vec{n})
\dot{\lambda}(a) dt
\ , 
\label{efun1}
\end{equation}
where the dot stands for a time derivative with respect to
the cosmological time t.

The following Fourier expansion
\begin{equation} 
\phi(\vec{x},t) = \frac {1} {(2 \pi)^{3/2}}
\int d^{3}k e^{-i \vec{k} \cdot \vec{x}}
\phi_{\vec {k}} (t)  
\end{equation}
can be combined with Eqs. (\ref{ng}) and (\ref{efun1}) to get:
\begin{eqnarray}
& &\Delta (\vec{x}_{_{P}},\vec{n}) = \nonumber \\  
\lefteqn{\frac {2i} {(2\pi)^{3/2}}\int_{t_{50}}^{t_{o}}
\dot{\lambda}(t) dt
\int d^{3}k e^{-i \vec{k} \cdot \vec{x}_{_{P}}} \phi_{\vec {k}} (t)
(\vec{k} \cdot \vec{n}) e^{-i \lambda (t)(\vec{k} \cdot \vec{n})}} 
\ , \nonumber \\
\label{f1}
\end{eqnarray}
where $\vec {k}$ is the 
comoving wavevector.
Using: (i) Equation (\ref{f1}), (ii) the expansion in 
spherical harmonics  
\begin{equation} 
\Delta (\vec{x}_{_{P}},\vec{n}) = 
\sum^{\infty}_{\ell=0} \sum_{m=-\ell}^{+\ell}
a_{\ell m}(\vec {x}_{_{P}})
Y_{\ell m}(\vec{n})   \ ,
\end{equation} 
(iii)
the identity
\begin{eqnarray}
& &(\vec{k} \cdot \vec{n}) e^{-i \lambda (t)(\vec{k} \cdot 
\vec{n})} =\nonumber \\
& & 4 \pi k \sum^{\infty}_{\ell=0} 
i^{\ell + 1} j^{\prime}_{\ell}(\lambda k)
\sum_{m=-\ell}^{+\ell} Y_{\ell m}(\vec{n}) Y_{\ell m}(\vec{k} /k) ,
\end{eqnarray}
where $j^{\prime}_{\ell}(x) = (d/dx) j_{\ell}(x)$, (iv)
the orthonormality relation for the spherical harmonics and (v)
the Fourier counterpart of
Eq. (\ref{lapla}), which reads
$\phi_{\vec{k}} = - (D \delta_{\vec{k}})/(ak^{2})$, where
$D = \frac {3}{2}(1 - \Omega_{\Lambda})H_{0}^{2}a_{0}^{3}$,
the following basic equations are easily obtained: 
\begin{equation}
a_{\ell m}(\vec {x}_{_{P}})= \frac {8 \pi D}{(2 \pi )^{3/2}} i^{\ell}
\int d^{3}k e^{-i \vec{k} \cdot \vec{x}_{_{P}}} k^{-2}
Y_{\ell m}(\vec{k} /k) F_{\ell}(\vec{k})
\label{sh1}
\end{equation}
where 
\begin{equation}
F_{\ell}(\vec{k}) = \int_{w_{50}}^{w_{o}} 
f(\vec{k},w) j^{\prime}_{\ell}(w) dw \ ,
\label{sh2}
\end{equation}
$w = k \lambda(a)$, and $f(\vec{k},w)$ denotes the function  
$a^{-1} \delta_{\vec{k}}(a)$ written in terms of the variable $w$.
Eqs. (\ref{sh1}) and (\ref{sh2}) have been obtained 
using the Fourier transform in all the space,
we have not introduced either boxes or 
N-body simulations yet. These 
elements will be only required in order to numerically
solve the basic equations (\ref{sh1}) and (\ref{sh2}).

Evidently, 
equation (\ref{sh1}) can be seen as a Fourier transform 
from the
$\vec{k}$ momentum space to the $\vec {x}_{_{P}}$ physical space,
and the function to be transformed
depends on function $F_{\ell}(\vec{k})$ given by 
Eq. (\ref{sh2}) and other well known
functions. In the physical space, each point is 
the position of a CMB observer who expand the temperature
contrast in spherical harmonics to get 
the coefficients $a_{\ell m}(\vec {x}_{_{P}})$.
From a mathematical point of view (to numerically 
compute $a_{\ell m}(\vec {x}_{_{P}})$), we can introduce 
an appropriate box in physical space with a large enough size 
(periodic universe).
There is another box in momentum space associated to
our $\vec {x}_{_{P}}$ box. In each node of the $\vec{k}$ box, 
the function 
$\delta_{\vec{k}}(a)$ must be evaluated a certain
number of times 
to perform 
the time integral involved in Eq. (\ref{sh2}).
As it is shown below, 
this integral can be performed using only
the $\delta_{\vec{k}}(a)$ values calculated
at each time step by 
the N-body simulation (see below for details).
According
to our arguments in Section 2, boxes of $128$ Mpc
size containing $128^{3}$ observers are appropriate 
because (a) the resulting function $\delta_{\vec{k}}(a)$
has all the necessary information about non-linearity and (b)
periodicity is not expected to be relevant.

\begin{figure}
\begin{centering}
\psfig{file=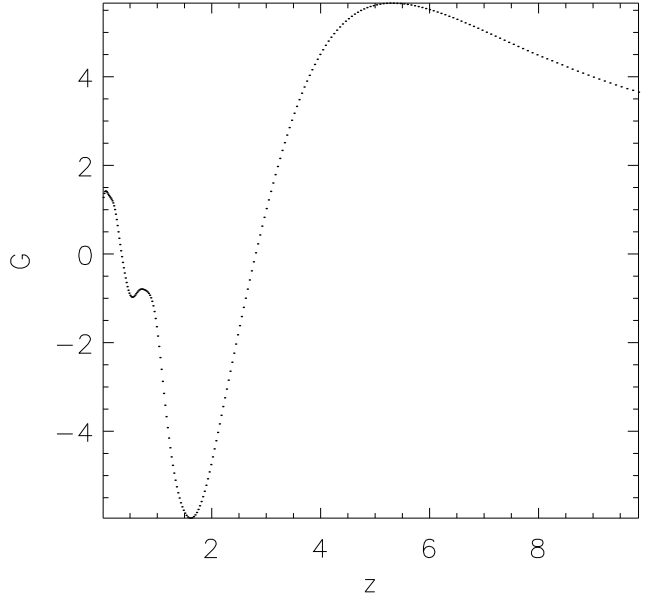,width=8.4cm}
\end{centering}
\caption{Function $G(z) = f(\vec{k}_{0},w(z))$ 
(see text)
for an certain node $\vec {k}_{0}$ of the computational box. 
Each point corresponds to a time step
of the N-body simulation.}
\label{fig-lx-error}
\end{figure}

Quantities $a_{\ell m}(\vec{x}_{_{P}})$ are evaluated 
on the nodes, $\vec{x}_{_{P}}$, of the assumed 
3D grid as follows: (i) At each time step, $t_{i}$, of the
N-body simulations (see above),
quantity $\delta_{\vec{k}}(a)$ is calculated
on the nodes of our $128^{3}$
cubic grid (in momentum space). 
At a given time $t_{i}$, only the quantities  
$f(\vec{k},w_{i-1})$ and $f(\vec{k},w_{i})$ 
are stored,
(ii) the integral in Eq. (\ref{sh2}) is
solved for every node 
and in each time interval [$t_{i-1}$,$t_{i}$] as it is described 
below and, 
then, results from all the time intervals are added to obtain 
$F_{\ell}(\vec {k})$. 
It has been verified that, inside any of the intervals ($t_{i-1}$, $t_{i}$), 
function $f_{i}(\vec{k},w)$ can be very well approximated by a
straight line and, as a consequence
of this fact, the integral in Eq. (\ref{sh2}) can be 
analytically calculated between $t_{i-1}$ and $t_{i}$.
The linearity of $f_{i}(\vec{k},w)$
is due to the fact that the time step of the N-body 
simulation is short enough; in order to display this
linearity, we present 
Fig. 2, which shows function $G(z) = f(\vec{k}_{0},w(z))$ for a
particular
node $\vec{k}_{0}$. Each point is a time step in the 
N-body simulation and
the linear behaviour between contiguous points is
evident. The same has been verified for many nodes.
From Fig. 2 it follows that function $f_{i}(\vec{k},w)$
also has a linear behaviour
between more separated points, 
this fact could be used in future to reduce
CPU time; nevertheless, in this paper, linearization 
is assumed between neighbouring points.
The equation of the line joining two neighbouring points 
\begin{equation} 
f_{i}(\vec{k},w) = w b_{i}(\vec{k}) + c_{i}(\vec{k}) 
\label{inter}
\end{equation}
involves
two coefficients to be computed from  the two stored 
values 
$f(\vec{k},w_{i-1})$ and $f(\vec{k},w_{i})$.
Then, using Eq (\ref{inter}), the 
value of the
integral (\ref{sh2}) can be analytically found 
in any interval [$t_{i-1}$,$t_{i}$], the result
is
\begin{eqnarray}
& &F_{\ell}(\vec{k})=\nonumber \\
\lefteqn{\sum_{i}\left[ w_{i} b_{i}(\vec{k}) + c_{i}(\vec{k})
\right] j_{\ell}(w_{i}) -      
\left[ w_{i-1} b_{i}(\vec{k}) + c_{i}(\vec{k})
\right] j_{\ell}(w_{i-1})}  \nonumber \\
\lefteqn{-b_{i}(\vec{k}) \int_{w_{i-1}}^{w_{i}} j_{\ell}(w) dw \ ;}
\label{clave}
\end{eqnarray}
it is noticeable that this formula only involves: the coefficients
appearing in Eq. (\ref{inter}), the spherical bessel functions, and
the integral of these functions appearing in the last term,
and (iii) after computing $F_{\ell}(\vec{k})$, Eq. (\ref{sh1}) --a Fourier
transform-- is used to 
obtain $a_{\ell m}(\vec {x}_{_{P}})$ on the
$128^{3}$ nodes of the $\vec {x}_{_{P}}$ grid. 
Thus, the expansion in spherical harmonics
of the CMB temperature contrasts corresponding to 
$128^{3}$ observers is obtained and, afterwards,
these data are used to perform averages.

Taking into account that $a_{400,0}$ is real, 
we calculate the i-averages 
$M_{400}^{i} = \langle a_{400, 0} \rangle$,
$C_{400}^{i} = \langle a_{400, 0}^{2} \rangle$,
$B_{400}^{i} = \langle a_{400, 0}^{3} \rangle$, and 
$K_{400}^{i} = \langle a_{400, 0}^{4} \rangle$ 
over the $128^{3}$ nodes of the $i$-th simulation.
Since these averages have been performed on a 
limited sample of rather close observers located in 
a particular region of the universe (nodes in our grid),
they appear to change from simulation to
simulation.
In order to calculate the statistical moments 
of the CMB, the simulations must be repeated a large enough 
number of times (to have a big enough sample of CMB observers
located in independent regions).
Then, the i-averages 
corresponding to $n$ different simulations ($i=1,2,...,n$)
must be averaged again to get the quantities
$M^{n}_{400}$, $C^{n}_{400}$, $B^{n}_{400}$, and $K^{n}_{400}$,
which must tend to 
the required CMB moments 
$M_{400}$, $C_{400}$, $B_{400}$, and $K_{400}$, respectively, 
as n tends to infinity.

\section{DISCUSSION AND RESULTS}

In this paper, a new method has been designed with
the essential aim of calculating deviations from Gaussianity 
due to the Rees-Sciama effect.
Our method is different from other ones previously
used to estimate deviations from Gaussianity due to 
weak lensing. According to White \& Hu (2001), there 
are too much scales contributing to lensing and {\em
simulating the full range of scales implied is currently 
a practical impossibility};
sophisticated techniques as "special 
projections" (Jain, Seljak \& White 2000) or
"tiling" (White \& Hu 2001) have been designed to circumvent 
the problem; nevertheless,
we are not considering lensing here and, as discussed
in Section 2, the estimate of the deviations with respect
to Gaussianity produced by the Rees-Sciama effect 
only requires simulations involving a moderated range
of scales, which can be performed (see Section 2).
In addition, in the Rees-Sciama case, some
analytical calculations (Section 3) have given formulae to 
simultaneously calculate the $a_{\ell m}(\vec {x}_{_{P}})$
quantities of many observers, from which, the 
successive statistical moments are calculated by performing
appropriate averages.

Condition (\ref{inter}) plays a fundamental role
in our numerical method. It leads to Eq. (\ref{clave}), which allows us
to compute $F_{\ell}(\vec {k})$ using a few stored data from the N-body 
simulation
and the spherical Bessel function $j_{\ell}$.
This computation of $F_{\ell}(\vec {k})$ is 
much faster than any other procedure based on the direct numerical
calculation of the integral in  
Eq. (\ref{sh2}). In spite of this fact, the quantities involved in 
Eq. (\ref{clave}) are (in this paper) evaluated step by step at each node 
of the computational grid and, consequently, the calculation 
of $a_{\ell m}(\vec {x}_{_{P}})$ --although feasible in modern
computers-- has not a low computational cost. 
It is worthwhile to remark that, once the
quantities $a_{\ell m}(\vec {x}_{_{P}})$ have been calculated,
we can obtain, not only one, but many different moments of the temperature 
distribution with very small
effort (by computing the necessary averages on nodes and 
realizations).

With the essential aim of finding non-Gaussian features and 
proving that the proposed 
method works, we have only considered the case $\ell=400$,
$m=0$. 
For very different $\ell$ values computations would be similar,
but the range of spatial scales, the box, and the resolution should
be chosen again taking into account Fig. 1.
The larger the $\ell$ value, the greater the necessary resolution.
More systematic estimations for other      
$\ell$ values --computationally expensive-- are in progress.
According to some comments in Section 1, the $m$
dependence can be obtained from the results of the m=0 case.

We have performed 90 simulations, in which,
the value of $\mid M^{i}_{400} \mid $ 
is typically a few times $10^{-19}$, whereas about $70 \% $ ($25 \% $)
of the absolute values of
$ a_{400,0} $ are of the order of $10^{-9}$ ($10^{-10}$). 
This means that --after ensuring the required numerical precision--
the positive and negative values of
$a_{400,0}$
strongly cancel among them to give a
$\mid M^{i}_{400} \mid $ value as 
small as reported above.
On the contrary, the positive and negative values of $a_{400,0}^{3}$
(most of them with an absolute value between orders 
$10^{-27}$ and $10^{-30}$)
do not undergo such a strong cancellation, and
a significant bispectrum appears (see below).

\begin{figure}
\begin{centering}
\psfig{file=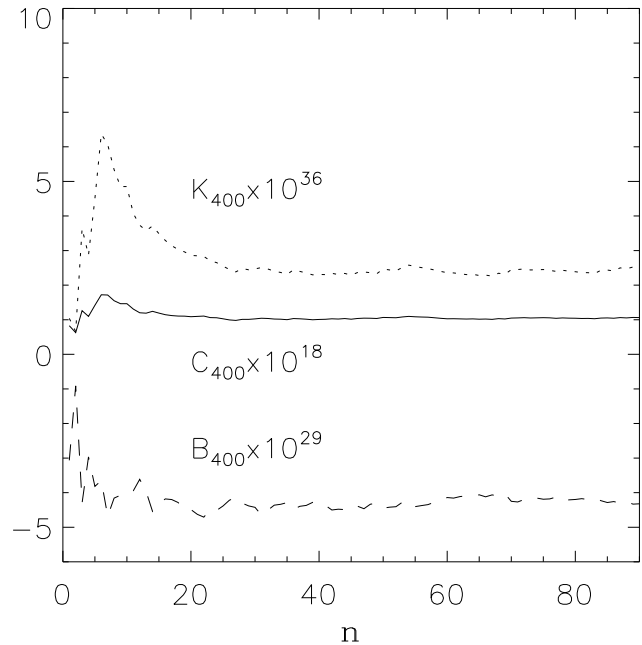,width=8.4cm}
\end{centering}
\caption{Quantities 
$C^{n}_{400} \times 10^{18}$, $B^{n}_{400} \times 10^{29}$, 
and $K^{n}_{400} \times 10^{36}$ 
vs. the number of simulations $n$.}
\label{fig-lx-error}
\end{figure}

Quantities
$C^{n}_{400}$, $B^{n}_{400}$, and $K^{n}_{400}$,
are displayed in Fig. 3 (in terms of 
the variable $n$, from $n=3$ to
$n=90$). As the number of simulations $n$ increases, 
the values of these quantities should approach
the cosmological limit values $C_{400}$, $B_{400}$, 
and $K_{400}$, respectively. 
Of course, the three curves in 
this panel seem to approach constant values as $n$
increases. In order to perform a quantitative study of
this behaviour, we have considered the n-interval [20,90], in which 
all the curves of Fig. 3 seem to approach their 
limits and, in this interval, we have computed
the means and variances of the above n-quantities. Thus,
we have found that, within
$2 \sigma $ confidence level, the following relations are satisfied:
$C_{400} \times 10^{18} = 1.04 \pm 0.05$, 
$B_{400} \times 10^{29} = -4.32 \pm 0.27$, 
$K_{400} \times 10^{36} = 2.45 \pm 0.26$.
The mean and variance 
(from $n=20$ to $n=90$) of the quantity 
$\Gamma^{n}_{400} = 3(C^{n}_{400})^{2}/K^{n}_{400}$
lead to 
the following relation  
$\Gamma_{400} = 1.34 \pm 0.09$, which is valid within 
$2 \sigma $ confidence level;
therefore, for $\ell = 400$ and $m=0$, a non-vanishing component of the
bispectrum and a non-Gaussian value of the ratio $\Gamma_{400}$ have
been found. It is noticeable the fact that quantities 
$C^{n}_{400}$, $B^{n}_{400}$, and $K^{n}_{400}$ really
converge as $n$ increases. There are no uncontrolled
errors producing strong oscillations and avoiding 
convergence, this verification can be seen as a severe test 
which has been successfully passed by the
proposed computational method.

The Rees-Sciama 
effect was studied by Seljak (1996) in 
the standard CDM model ($\Lambda = 0$). Four cases corresponding
to different values of the parameters $\sigma_{8}$ 
and $\Omega_{m0}h^{2}$ were considered by this author, who 
obtained $C_{400}$ values between $10^{-17}$ and $10^{-18}$
corresponding to a signal, $(\Delta T /T)_{rms}$, between
$10^{-6}$ and $10^{-7}$. Although we have not computed 
all the $C_{\ell}$ quantities, the resulting $C_{400}$ 
is comparable to the minimum values obtained by Seljak;
hence, the non-linear gravitational effect should be of the order
of $10^{-7}$, which is below the observational
sensitivities of current and planned experiments.

The model of structure formation assumed here
is suggested by far supernovae and
recent CMB observations. However, other models cannot be 
ruled out yet. Regardless the structure formation model,
our method is ready to do estimations. In any case, 
our results are challenging to understand the data
in forthcoming CMB experiments

\vspace{0.5 cm}

\noindent
{\bf ACKNOWLEDGMENTS}. This work has been partially 
supported by the Spanish MCyT 
(project AYA2000-2045). One of us (VQ)
is Marie Curie Research Fellow from the EU (grant HPMF-CT-1999-00052).
Calculations
were carried out on a SGI Origin 2000s at the Centro de Inform\'atica
de la Universitat de Val\`encia.

\vspace{0.5 cm}

\noindent
{\large{\bf REFERENCES}}\\ 
\noindent
Eke V., Cole S., Frenk C.S., 1996, MNRAS, 282, 263\\ 
Fullana M.J., S\'aez D., 2000, New Astronomy, 5, 109\\
Grishchuck L.P., Martin J., 1997, Phys. Rev., 56D, 1924\\  
Hockney R.W., Eastwood J.W., 1988, Computer Simulations
Using Particles (Bristol: IOP Publishings)\\
Hu W., The Angular Trispectrum of the CMB, astro-ph/0105117\\
Jain B., Seljak U., White S., 2000, ApJ, 530, 547\\
Mart\'{\i}nez-Gonz\'alez E., Sanz J.L., Silk J., 1990, ApJ, 355, L5\\
Mart\'{\i}nez-Gonz\'alez E., Sanz J.L., Silk J., 1994, ApJ, 436, 1\\
Mollerach S., Gangui A., Lucchin F., Matarrese S., 1995, ApJ, 453, 1\\
Munshi D., Souradeep T., Starobinski A. A., 1995, ApJ, 454, 552\\ 
Quilis V., Ib\'a\~{n}ez J.M., S\'aez D., 1998, ApJ, 502, 518\\
Sanz J.L., Mart\'{\i}nez-Gonz\'alez, E., Cay\'on L., Silk J.L., 
Sugiyama N., 1996, ApJ, 467, 485\\
Seljak U., 1996, ApJ, 460, 549\\ 
Tuluie R., Laguna P., Anninos P., 1996, ApJ, 463, 15\\  
White M., Hu W., 2000, ApJ, 537, 1\\ 

\end{document}